\documentclass[11pt]{article}
\usepackage{amsfonts}
\usepackage{amsmath,amssymb}
\usepackage{epsfig}
\def\dref#1{(\ref{#1})}

\begin{document}

\begin{center}
{{\Large \bf Cost and Effects of Pinning Control for Network
Synchronization}}
 \footnote{
{\footnotesize This work is supported by the National
 Science Foundation of
 China under grants 60674093, 60334030.}}
\end{center}
\vskip 0.2cm
\begin{center}
 Rong Li $^{\textrm{a}}$ \footnote{ {\footnotesize Corresponding
author: lirong@pku.edu.cn}},
  \quad Zhisheng Duan $^{\textrm{a}}$,
  \quad Guanrong Chen $^{\textrm{b}}$
 \end{center}
\begin{center}
{\small \it  $^{\textrm{a}}$ Department of Mechanics and Aerospace
Technology, Peking University, Beijing, P. R. China
\\ $^{\textrm{b}}$ Department of Electronic Engineering, City University of Hong Kong, Hong Kong}
\end{center}

\vspace*{1\baselineskip}
\begin{center}
{\begin{minipage}{127mm} {\small {\bf Abstract.} In this paper, the
problem of pinning control for synchronization of complex dynamical
networks is discussed. A cost function of the controlled network is
defined by the feedback gain and the coupling strength of the
network. An interesting result is that lower cost is achieved by the
control scheme of pinning nodes with smaller degrees. Some rigorous
mathematical analysis is presented for achieving lower cost in the
synchronization of different star-shaped networks. Numerical
simulations on some non-regular complex networks generated by the
Barab\'{a}si-Albert model and various star-shaped networks are shown
for verification and illustration.}

\vspace*{0.5\baselineskip} {{\bf Keywords:}
 complex dynamical network, pinning control, exponential stability.}

\end{minipage}
}
\end{center}

\abovedisplayskip=0.13cm \abovedisplayshortskip=0.06cm
\belowdisplayskip=0.13cm \belowdisplayshortskip=0.08cm

\def\dis{\displaystyle}


\section{Introduction and problem formulation}
\quad Complex networks are currently being studied across many
fields of sciences, including physics, chemistry, biology,
mathematics, sociology and engineering
\cite{r2,r12,r3,r5,r4,r11,r6,r1}. A complex network is a large set
of interconnected nodes, in which a node is a fundamental unit with
specific contents. Examples of complex networks include the
Internet, food webs, cellular neural networks, biological neural
networks, electrical power grids, telephone cell graphs, etc.
Recently, synchronization of complex networks of dynamical systems
has received a great deal of attention from the nonlinear dynamics
community \cite{jia54,r7,jia51,jia52,jia56}. A special control
strategy called \textit{pinning control} is used to achieve
synchronization of complex networks; that is, only a fraction of the
nodes or even a single node is controlled over the whole network
\cite{jia53,r8,jia55,jia57}. This control method has become a common
technique for control, stabilization and synchronization of coupled
dynamical systems. In general, different nodes have different
degrees in a network, thus a natural question is how different the
effect would be when nodes with different degrees are pinned.

Consider a dynamical network consisting of N identical and
diffusively coupled nodes, with each node being an $n$-dimensional
dynamical system. The state equations of the network are
\begin{equation} \label{syn1}
\dot{x}_i(t)=f(x_i(t),t)+c\sum ^{N}_{j=1}a_{ij}\Gamma x_j(t),\qquad
i=1,2,\cdots,N,
\end{equation}
where $f(\cdot)$ is the dynamical function of an isolated node,
$x_i=(x_{i1},x_{i2},\cdots,x_{in})\in R^{n}$ are the state variables
of node $i$, constant $c>0$ represents the coupling strength, and
$\Gamma\in R^{N\times N}$ is the inner linking matrix. Moreover, the
coupling matrix $A=(a_{ij})\in R^{N\times N}$ represents the
coupling configuration of the network: If there is a connection
between node $i$ and node $j$ $(i\neq j)$, then $a_{ij}=a_{ji}=1$;
otherwise, $a_{ij}=a_{ji}=0$ $(i\neq j)$; the diagonal entries of
$A$ are defined by
\begin{equation} \label{syn2}
a_{ii}=-\sum ^{N}_{j=1 \atop j\neq i}a_{ij},\qquad i=1,2,\cdots,N.
\end{equation}
Suppose that the network is connected in the sense of having no
isolated clusters. Then, the coupling matrix $A$ is irreducible.
From Lemma 2 of \cite{jia56}, it can be proved that zero is an
eigenvalue of $A$ with multiplicity one and all the other
eigenvalues of $A$ are strictly negative.

Network (\ref{syn1}) is said to achieve (asymptotical)
synchronization if
\begin{equation} \label{syn3}
x_1(t)\rightarrow
x_2(t)\rightarrow\cdots\rightarrow x_N(t)\rightarrow s(t),\qquad
\textrm{as}\quad t\rightarrow\infty,
\end{equation}
where, because of the diffusive coupling configuration, $s(t)$ is a
solution of an isolated node, which can be an equilibrium, a
periodic or a chaotic orbit. As shown in
\cite{jia53,r8,jia55,jia57}, this can be achieved by controlling
several nodes (or even only one node) of the network. Without loss
of generality, suppose that the controllers are added on the last
$N-k$ nodes of the network, so that the equations of the controlled
network can be written as
\begin{equation} \label{cost7}
 \begin{array}{l}
 \dot x_i(t)=f(x_i(t),t)+c\sum _{j=1}^{N}a_{ij}\Gamma
 x_j(t),\qquad\qquad\qquad\qquad\qquad\;
i=1,2,\cdots,k,\\
\\
 \dot x_i(t)=f(x_i(t),t)+c\sum_{j=1}^{N}a_{ij}\Gamma x_j(t)-c\varepsilon_i\Gamma(x_i(t)-s(t)),\qquad
i=k+1,k+2,\cdots,N,\end{array}
\end{equation}
where the feedback gains $\varepsilon_i$ are positive constants. It
can be seen that synchronizing all states $x_i(t)$ to $s(t)$ is
determined by the dynamics of an isolated node, the coupling
strength $c>0$, the inner linking matrix $\Gamma$, the feedback
gains $\varepsilon_i\geq0$, and the coupling matrix $A$.

As discussed in \cite{r8,jia55,jia57}, to achieve synchronization of
complex dynamical networks, the controllers are generally preferred
to be added to the nodes with larger degrees. However, it is also
known that, to achieve a certain synchronizability of the network,
the feedback gains $\varepsilon_i$ usually have to be quite large.
In \cite{jia53}, when a single controller is used, the coupling
strength $c$ has to be quite large in general. From the view point
of realistic applications, these are not expected and sometimes
cannot be realized. Practically, a designed control strategy is
expected to be effective and also easily implementable. In this
paper, for various star-shaped networks and non-regular complex
networks, a new concept of cost function is introduced to evaluate
the efficiency of the designed controllers. It is found that
surprisingly the cost can be much lower by controlling nodes with
smaller degrees than controlling nodes with larger degrees. As will
be seen, moreover, both the feedback gains $\varepsilon_i$ and the
coupling strength $c$ can be much smaller than those used in
\cite{jia53,r8,jia55,jia57}.

The outline of this paper is as follows. In Section 2, a new
definition of cost function and some mathematical preliminaries are
given. Stability of different star-shaped networks controlled by
pinning some nodes with small degrees are analyzed in Sections 3 and
4, respectively, where some simulated examples of dynamical networks
are compared for illustration and verification. In Section 5,
pinning control of non-regular complex dynamical networks of chaotic
oscillators is studied through numerical simulations. Finally,
Section 6 concludes the paper.

\section{The cost of pinning control}

\quad Denote $e_i(t)=x_i(t)-s(t)$, where $s(t)$ satisfies $\dot
s(t)=f(s(t))$. Then, the error equations of network \dref{syn1} can
be written as
$$
\dot e_i(t)=f(x_i(t),t)-f(s(t))+c\sum ^{N}_{j=1}a_{ij}\Gamma
 e_j(t),\qquad\qquad\qquad\;\;
i=1,2,\cdots,N,
$$
while the error equations of the controlled network \dref{cost7} can
be written as
\begin{equation}\label{syn5}
\dot e_i(t)=f(x_i(t),t)-f(s(t))+c\sum ^{N}_{j=1}\tilde{a}_{ij}\Gamma
 e_j(t),\qquad\qquad\qquad\;\;
i=1,2,\cdots,N,
\end{equation}
where
$\tilde{a}_{ii}=a_{ii}-\varepsilon_i,\,\varepsilon_i>0,\,i=k+1,k+2,\cdots,N$,
and $\tilde{a}_{ij}=a_{ij}$ otherwise. Let
$\tilde{A}=(\tilde{a}_{ij})\in R^{N\times N}$, and denote
$e(t)=(e_1(t),e_2(t),\cdots,e_N(t))^T$.

Differentiating \dref{syn5} along $s(t)$ gives
\begin{equation}\label{syn6}
\dot e(t)=D(f(s(t)))e(t)+c\Gamma e(t)\tilde{A}^T.
\end{equation}
By analyzing the matrix $\tilde{A}^T$, it is easy to see that all
the eigenvalues of $\tilde{A}^T$ are negative, which are denoted by
$$
0>\lambda_1\geq\cdots\geq\lambda_N.
$$
There exists an orthogonal matrix $U$ such that
$\tilde{A}^T=UJU^{-1}$, where
$J=diag\{\lambda_1,\lambda_2,\cdots,\lambda_N\}$. Let
$\tilde{e}(t)=e(t)U$. Then, from \dref{syn6}, one has
\begin{equation}\label{syn7}
\dot
{\tilde{e}}_i(t)=[Df(s(t))+c\lambda_i\Gamma]\tilde{e}_i(t),\qquad\qquad\qquad\;\;
i=1,\cdots,N.
\end{equation}
Therefore, the local stability problem of network \dref{cost7} is
converted into the stability problem of the $N$ independent linear
systems \dref{syn7}. When the system function of an isolated node
and the inner linking matrix $\Gamma$ are fixed, the stability
problem of systems \dref{syn7} are dependent on the coupling
strength $c$ and the eigenvalues of $\tilde{A}$. Clearly, the
smaller the eigenvalues of the matrix $\tilde{A}$ are
($\lambda_i<0,\,i=1,\cdots,N$), the smaller the coupling strength
$c>0$ is needed to guarantee the same synchronizability of the
network \dref{cost7}, if systems \dref{syn7} have unbounded
synchronization regions \cite{jia52}. In the following, a cost
function is introduced to describe the efficiency of the
controllers.

\textbf {Definition 1} (Cost Function) Suppose that the feedback
gain matrix is $G=diag\{\varepsilon_1,\cdots,$ $\varepsilon_N\}$,
where $\varepsilon_i\geq0,\,i=1,\cdots,N$, are given as in
\dref{cost7}. The Cost Function is defined as
$$CF=c\sum_{i=1}^{N} \varepsilon_i.$$

\textbf {Remark 1} The smaller the $CF$, the more efficient a
control strategy to achieve the same goal of control, and the easier
to be implemented.

In order to discuss the effects of pinning control, the following
Lemmas are needed.

\textbf {Lemma 2} \cite{horn85} Let $A=[a_{ij}]\in C^{n\times n}$ be
Hermitian, and let $a_{nn}\leq\cdots\leq a_{22}\leq a_{11}$ be a
rearrangement of its diagonal entries in increasing order. Let the
eigenvalues of $A$ be ordered as
\begin{equation} \label{cost4}
\lambda_{\min}=\lambda_n\leq\lambda_{n-1}\leq\cdots\leq\lambda_2\leq\lambda_1=\lambda_{\max}.
\end{equation}
Then
$$\begin{array}{l}
\textrm{(i)}\quad\, \lambda_n\leq a_{ii}\leq\lambda_1 \quad
\textrm{for
all} \quad i=1,\cdots,n,\\
\textrm{(ii)}\quad a_{11}+a_{22}\leq \lambda_2,
\quad\textrm{if}\quad \lambda_1=0.
\end{array}
$$

\textbf {Remark 2} For diffusive networks, the smaller the
$\lambda_2$, the easier the synchronization, if the network has an
unbound synchronized region \cite{jia52}. However, Lemma 2 shows
that $\lambda_2$ is related to $a_{11}+a_{22}$, where $a_{11}$ and
$a_{22}$ are determined by two smallest nodes. Therefore, in this
case, in order to improve the synchronizability, these small nodes
should be pinned.

\textbf {Lemma 3} Let $A\in C^{n\times n}$ be Hermitian, and
$$ \label{syn4}
 \tilde{A}=\left[\begin{array}{cc}
 A_1  &  A_{12}  \\
 A_{21}^T    &  A_2+D
\end{array}\right],
$$
where $A_1\in R^{k\times k}$, $A_{12}\in R^{k\times(N-k)}$, $A_2\in
R^{(N-k)\times(N-k)}$ and
$D=diag\{\varepsilon_{k+1},\cdots,\varepsilon_N\}$. Then,
$\tilde{A}<-\alpha I,\,\alpha>0$, if and only if $A_1<-\alpha I_1$
and $A_2+D-A_{12}^T(A_1+\alpha I_1)^{-1}A_{12}<-\alpha I_2 $, where
$I_1$ and $I_2$ are identity matrixes with appropriate dimensions.

\textbf {Remark 3} Matrix $D$, i.e. $\varepsilon_i$ can be
determined by the LMI method \cite{Lmi}. Suppose the eigenvalues of
$A_1$ are $\beta_1\geq\beta_2\geq\cdots\geq\beta_k$. Then
$-\alpha>\beta_1$; and if $\varepsilon_i\rightarrow-\infty$, then
$-\alpha\rightarrow\beta_1.$

\section{Simple star-shaped networks}

\quad Now, consider a network consisting of $N$ identical nodes with
a simple star-shaped coupling configuration: there exists a central
node which connects to all the other non-central nodes, and there
are no direct connections among the non-central nodes. Fig. 1 gives
an example of such a network.
\begin{center}
\quad\unitlength=1cm \hbox{\hspace*{0.1cm} \epsfxsize6.6cm
\epsfysize6cm \epsffile{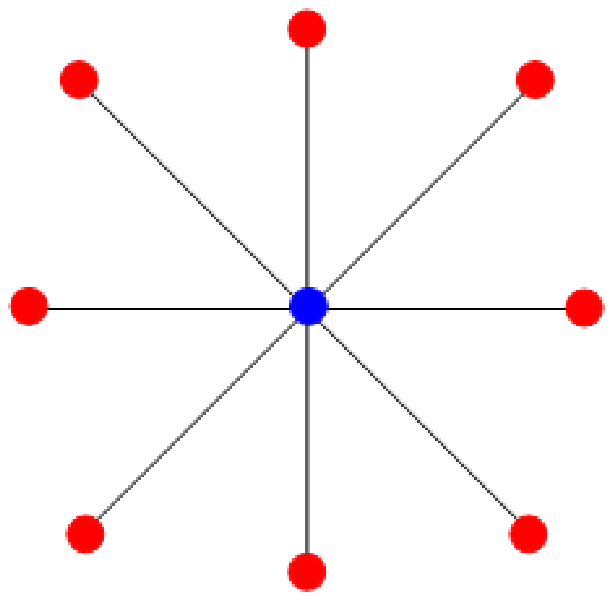}}
\end{center}
\vskip-0.6cm
\qquad\qquad\qquad\qquad\qquad\qquad\qquad\qquad\qquad\quad\, Fig.
1\,\,\,\, $N=9$

The coupling matrix $A$ of network \dref{syn1} with a simple
star-shaped coupling configuration can been written as
\begin{equation} \label{syn4}
 A=\left[\begin{array}{cc}
 A_1  &  A_{12}  \\
 A_{21}    &  A_2
\end{array}\right],
\end{equation}
where $A_1=-N+1$, $A_{12}=(1,1,\cdots,1)\in R^{(N-1)}$,
$A_{21}=A_{12}^T$, and $A_2=-I_{(N-1)\times(N-1)}.$

Let the feedback gain matrix be $
G=diag\{0,\varepsilon,\cdots,\varepsilon\}\in R^{N\times N};$ that
is, the controllers are added to all the non-central nodes for
simplicity in this discussion. Then, the coupling matrix of the
controlled network \dref{cost7} is
$$\tilde{A}=\left[\begin{array}{cc}
 A_1  &  A_{12}  \\
 A_{21}    &  A_2-\varepsilon I
\end{array}\right].
$$

\textbf {Theorem 1} If there are constants $\varepsilon>0$ and
$\tilde{k}>0$, such that $N-\tilde{k}-1>0$,
$\varepsilon>\tilde{k}-1$ and
$\varepsilon>\frac{\tilde{k}(N-\tilde{k})}{N-\tilde{k}-1}$, then
$\lambda_i(\tilde{A})<-\tilde{k},\,i=1,\cdots,N$, where
$\lambda_i(\tilde{A}),\,i=1,\cdots,N,$ are the eigenvalues of the
matrix $\tilde{A}$.

{\bf Proof:} Consider the matrix
$$
 \tilde{A}+\tilde{k}I=\left[\begin{array}{cc}
 A_1+\tilde{k} &  A_{12}  \\
 A_{21}  &  (-1-\varepsilon+\tilde{k})I
\end{array}\right].
$$
By the assumptions of the theorem, $(-1-\varepsilon+\tilde{k})I<0$.
Let
$\bar{A}=(A_1+\tilde{k}I)-A_{12}((-1-\varepsilon+\tilde{k})I)^{-1}A_{21}.$
Then
$$
 \begin{array}{ccl}
 \bar{A} & = & -N+1+\tilde{k}+\frac{N-1}{1+\varepsilon-\tilde{k}} \\
   & < & -N+1+\tilde{k}+\frac{N-\tilde{k}-1}{N-1}(N-1) \\
   & = & 0. \end{array}
$$
From the Schur complement Lemma, $\bar{A}+\tilde{k}I<0$, which leads
to the assertion of the theorem. \hfill $\square$

\textbf {Remark 4} Theorem 1 shows that under the control strategy
of adding controllers to all the non-central nodes, as the constant
$\varepsilon>0$ increases, the eigenvalues of the matrix $\tilde{A}$
will approach $-(N-1)$. Further, suppose that $N>2$ and take
$\tilde{k}=1$. Then, based on Theorem 1, to guarantee that
$\lambda_i(\tilde{A})<-1,\,i=1,\cdots,N$, the feedback
 gain $\varepsilon$ only needs to satisfy
\begin{equation}\label{cost2}
\varepsilon>\frac{N-1}{N-2}.
\end{equation}
In this case, if the coupling strength of network \dref{syn4} is
$c$, then the cost function is $CF=c\varepsilon(N-1)$. From
\dref{cost2} and through direct calculations, it is easy to see that
under the control strategy of adding controllers to the non-central
nodes, $CF>\frac{(N-1)^2}{(N-2)}$ suffices to guarantee that
$\lambda_i(\tilde{A})<-1,\,i=1,\cdots,N$.

\textbf {Remark 5} Consider the control strategy that a single
controller is added to the central node. In this case, the feedback
gain matrix is given by $G=diag\{\varepsilon,0,\cdots,0\}\in
R^{N\times N}$ and $CF=c\varepsilon$. Let $\lambda_1(\tilde{A})$
represent the largest eigenvalue of matrix $\tilde{A}$. Then, based
on Lemma 2, $\lambda_1(\tilde{A})\geq-1$ no matter how large the
$\varepsilon$, or equally the cost function $CF$, is taken. This
shows that in order to guarantee $\lambda_i$ be small, the small
nodes should be pinned.

Some simulations on the network \dref{syn4}, shown in Fig. 1, are
presented in Figs. 2$\sim$3, where the dynamics of an isolated node
is a chaotic oscillator \cite{r9}, $\Gamma=diag\{0,\,1,\,0\}$, and
the synchronized state is set to be $s(t)=[7.9373\;7.9373\;21]$,
which is an unstable equilibrium point of an isolated node
\cite{r10}.

\begin{center}
\quad \unitlength=1cm \hbox{\hspace*{0.1cm} \epsfxsize7.5cm
\epsfysize5.8cm \epsffile{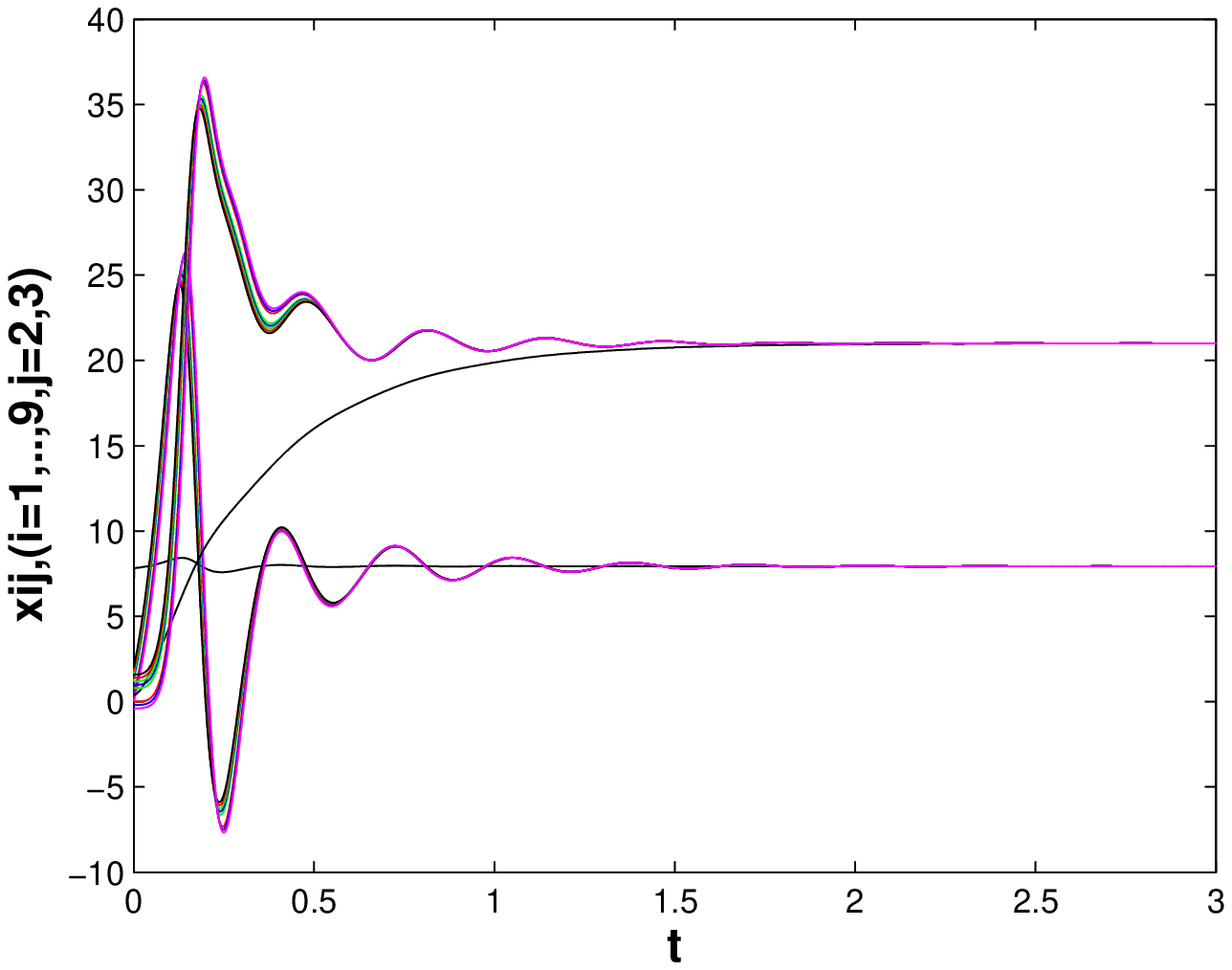}\qquad \epsfxsize7.5cm
\epsfysize5.8cm \epsffile{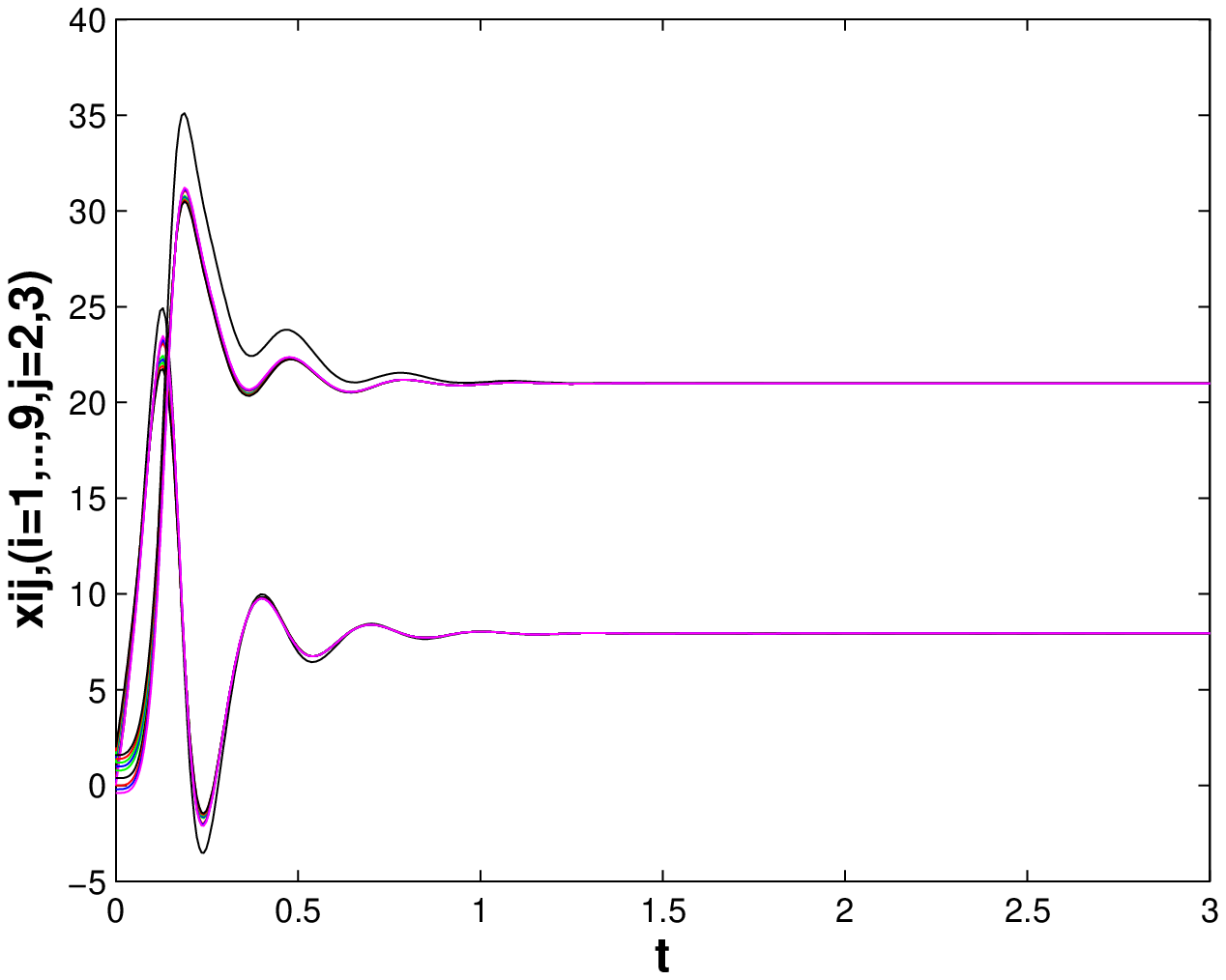}}
\end{center}
\vskip-0.6cm \qquad\qquad\qquad {\small (a)
$CF=c\varepsilon=300c=3000$. \qquad\qquad\qquad\qquad (b)
$CF=c\varepsilon(N-1)=1.5(N-1)c=120$.}
\vskip0.1cm \qquad {\small Fig. 2 \,\,\,\, Synchronization of the
simple star-shaped coupling network \dref{syn4} with coupling
strength $c=10$: (a) adding a single controller to the central node;
(b) adding controllers to all the non-central nodes.}
\begin{center}
\quad \unitlength=1cm \hbox{\hspace*{0.1cm} \epsfxsize7.5cm
\epsfysize5.8cm \epsffile{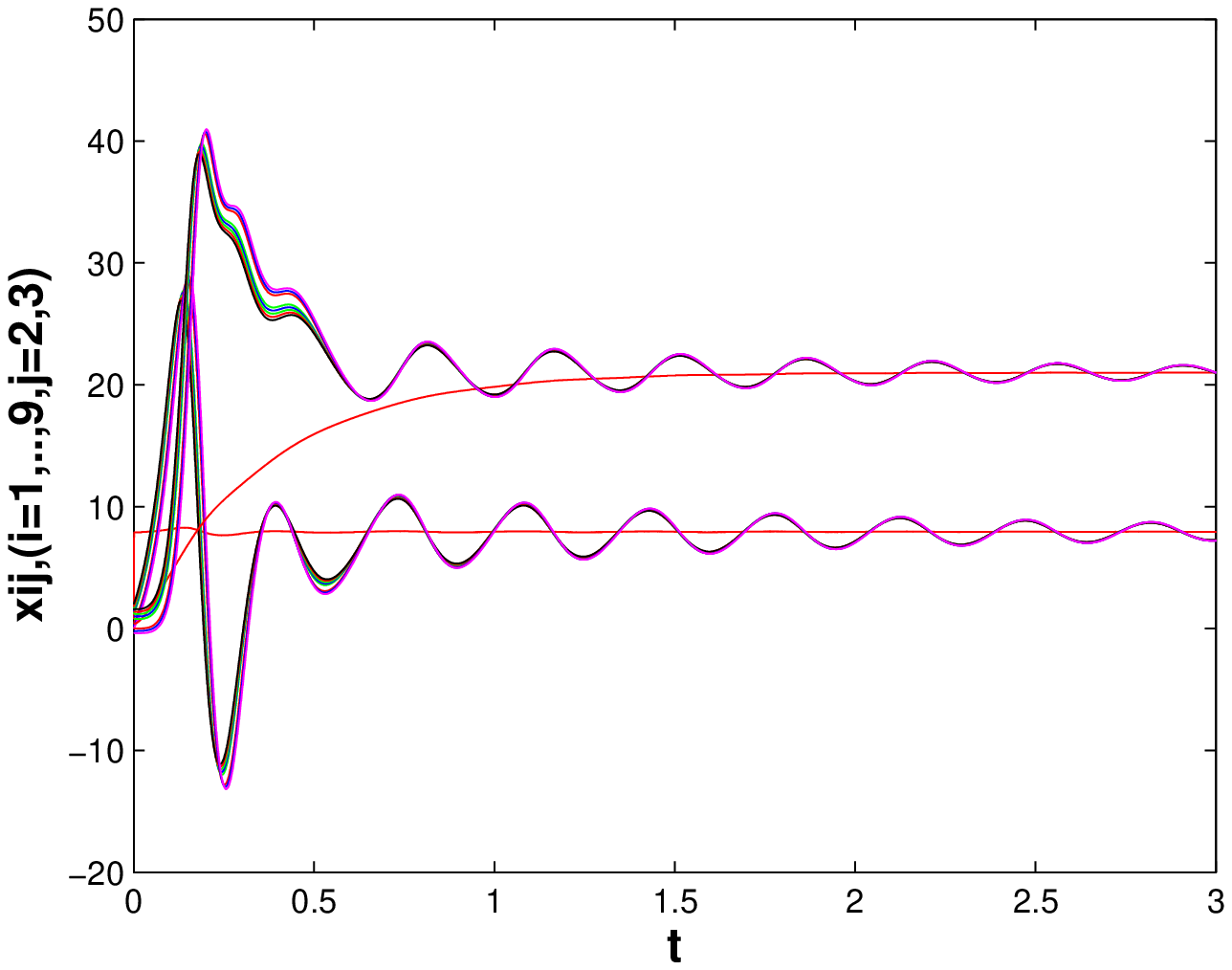}\qquad \epsfxsize7.5cm
\epsfysize5.8cm \epsffile{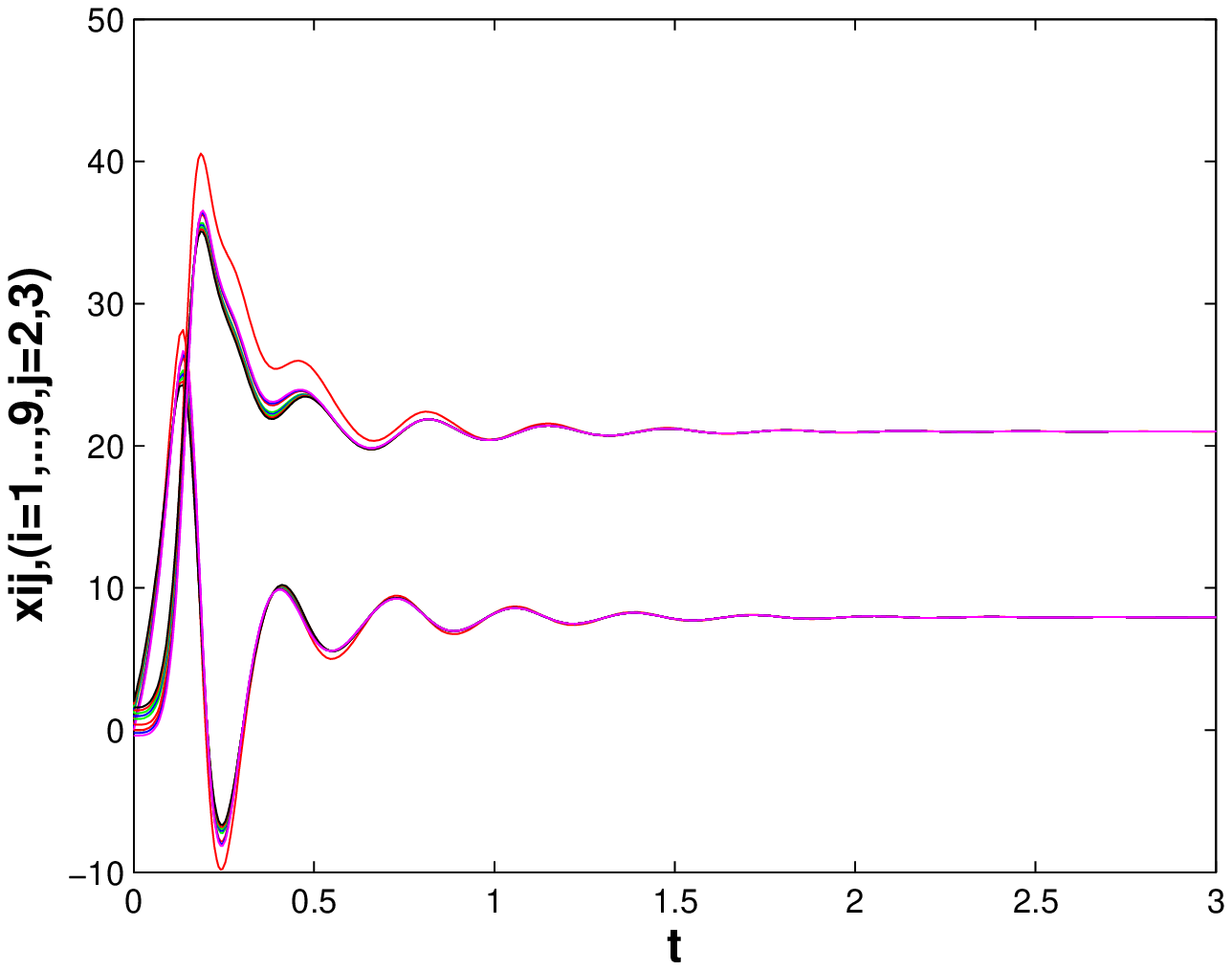}}
\end{center}
\vskip-0.6cm \qquad\qquad\qquad {\small (a)
$CF=c\varepsilon=500c=3500$. \qquad\qquad\qquad\qquad (b)
$CF=c\varepsilon(N-1)=1.5(N-1)c=84$.}
\vskip0.1cm \qquad {\small Fig. 3\,\,\,\, Synchronization of the
simple star-shaped coupling network \dref{syn4} with coupling
strength $c=7$: (a) pinning the central node; (b) pinning all the
non-central nodes.}

From Figs. 2$\sim$3, it can be seen that even with both a small
coupling strength $c$ and a small cost function $CF$, the
synchronization of network \dref{syn4} can be faster to achieve when
the controllers are added to all the non-central nodes (nodes with
smaller degrees) than the case that a single controller is added to
the central node (the node with a larger degree). Although the more
controllers are needed for pinning nodes with smaller degrees, the
total cost is still lower and the control effect is better in
comparison.

\section{Clusters of star-shaped networks with global coupling}

\quad Consider a network consisting of $N$ identical nodes in
clusters of star-shaped coupling configuration: there are $k$
central nodes which are connected to each other; each central node
may have different numbers of non-central nodes attached to it;
there are no direct connections among the non-central nodes; and any
central node has no direct connection to a non-central node attached
to another central node. Fig. 4 gives an example of such a network
for the case of $k=3$.
\begin{center}
\quad \unitlength=1cm \hbox{\hspace*{0.1cm} \epsfxsize9cm
\epsfysize6cm \epsffile{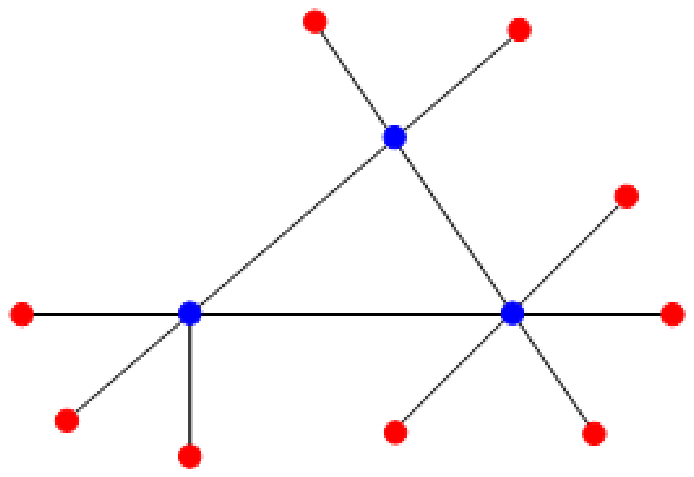}}
\end{center}
\vskip-0.6cm \qquad\qquad\qquad\qquad\qquad\quad Fig. 4\,\,\,\, A
cluster of star-shaped network with $k=3$

Without loss of generality, let the first $k$ nodes be central
nodes, each connecting to $n_i$ non-central nodes, where $n_i$
satisfy that $n_1\leq n_2\leq\cdots\leq n_k.$ Clearly,
$N=k+\sum_{i=1}^kn_i$. The coupling matrix $A$ of network
\dref{syn1} in such clusters of star-shaped coupling configuration
can been written as
\begin{equation} \label{cost6}
 A=\left[\begin{array}{cc}
 A_1  &  A_{12}  \\
 A_{21}  &  A_2
\end{array}\right],
\end{equation}
where $A_1\in R^{k\times k}$, $A_{12}\in R^{k\times(N-k)}$,
$A_{21}=A_{12}^T\in R^{(N-k)\times k}$, and $A_2\in
R^{(N-k)\times(N-k)}.$ Let $\alpha_i=(1,1,\cdots,1)\in
R^{n_i}$,\,\,$i=1,\cdots,k$. Then

$
 A_1=\left[\begin{array}{cccc}
 -k+1-n_1   &     1         & \cdots  & 1\\
     1      &  -k+1-n_2     & \cdots  & 1\\
  \vdots    &   \vdots      & \ddots  &\vdots\\
     1      &     1         & \cdots  &-k+1-n_k\end{array}\right]
$,\quad $
 A_{12}=\left[\begin{array}{cccc}
 \alpha_1 &           &          &        \\
          & \alpha_2  &          &        \\
          &           &  \ddots  &        \\
          &           &          & \alpha_k\end{array}\right],
$

$A_{21}=\left[\begin{array}{cccc}
 \alpha^T_1 &           &          &      \\
          & \alpha^T_2  &          &      \\
          &           &  \ddots  &     \\
          &           &          & \alpha^T_k\end{array}\right]
$,\quad $A_2=-I_{(N-k)}.$\\
Let the feedback gain matrix be
$
 G=\left[\begin{array}{cc}
 0  &  0  \\
 0  &  \varepsilon I_{(N-k)}
\end{array}\right].
$ Then, the corresponding coupling matrix $\tilde{A}$ of the
controlled network \dref{cost7} is
\begin{equation}\label{syn13}
\tilde{A}=A-G.
\end{equation}

\textbf {Theorem 2} If there are constants $\varepsilon>0$ and
$\tilde{k}>0$, such that $n_1>\tilde{k}$, $\varepsilon>\tilde{k}-1$
and $\varepsilon>\frac{\tilde{k}(n_1+1-\tilde{k})}{n_1-\tilde{k}}$,
then $\lambda_i(\tilde{A})<-\tilde{k},\,i=1,\cdots,N$.

{\bf Proof:} Consider the matrix
$$
 \tilde{A}+\tilde{k}I=\left[\begin{array}{cc}
 A_1+\tilde{k}I  &  A_{12}  \\
 A_{21}  &  (-1-\varepsilon+\tilde{k})I
\end{array}\right].
$$
By the assumptions of the theorem, $(-1-\varepsilon+\tilde{k})I<0$.
Let
$\bar{A}=(A_1+\tilde{k}I)-A_{12}((-1-\varepsilon+\tilde{k})I)^{-1}A_{21}$.
Then $\bar{A}$ can be expressed as
$$
 \bar{A}=\left[\begin{array}{cccc}
 -k+1-n_1+\tilde{k}+\frac{n_1}{1+\varepsilon-\tilde{k}}   &     1         & \cdots  & 1\\
     1      &  -k+1-n_2+\tilde{k}+\frac{n_2}{1+\varepsilon-\tilde{k}}     & \cdots  & 1\\
  \vdots    &   \vdots      & \ddots  &\vdots\\
     1      &     1         & \cdots
     &-k+1-n_k+\tilde{k}+\frac{n_k}{1+\varepsilon-\tilde{k}}\end{array}\right].
$$
Let $m_i=
-k+1-n_i+\tilde{k}+\frac{n_i}{1+\varepsilon-\tilde{k}}+(k-1)\cdot1,\quad
i=1,2,\cdots,k.$ Then
$$
 \begin{array}{ccl}
 m_i & = & -n_i+\tilde{k}+\frac{n_i}{1+\varepsilon-\tilde{k}} \\
   & < & -n_i+\tilde{k}+\frac{n_1-\tilde{k}}{n_1}n_i           \\
   & = & n_i(\frac{-\tilde{k}}{n_1}) +\tilde{k} \\
   & = & \tilde{k}(1-\frac{n_i}{n_1})\leq 0.\end{array}
$$
According to the Gerschgorin disc theorem \cite{horn85},
$\bar{A}<0$. By the Schur complement Lemma,
$\tilde{A}+\tilde{k}I<0$, which leads to the assertion of the
theorem. \hfill $\square$

Some simulations on the network \dref{cost6} shown in Fig. 4 are
presented in Fig. 5, where the dynamics of an isolated node, the
inner liking matrix $\Gamma$, and the synchronized state are the
same as that given in Section 2.
\begin{center}
\quad \unitlength=1cm \hbox{\hspace*{0.1cm} \epsfxsize7.5cm
\epsfysize5.8cm \epsffile{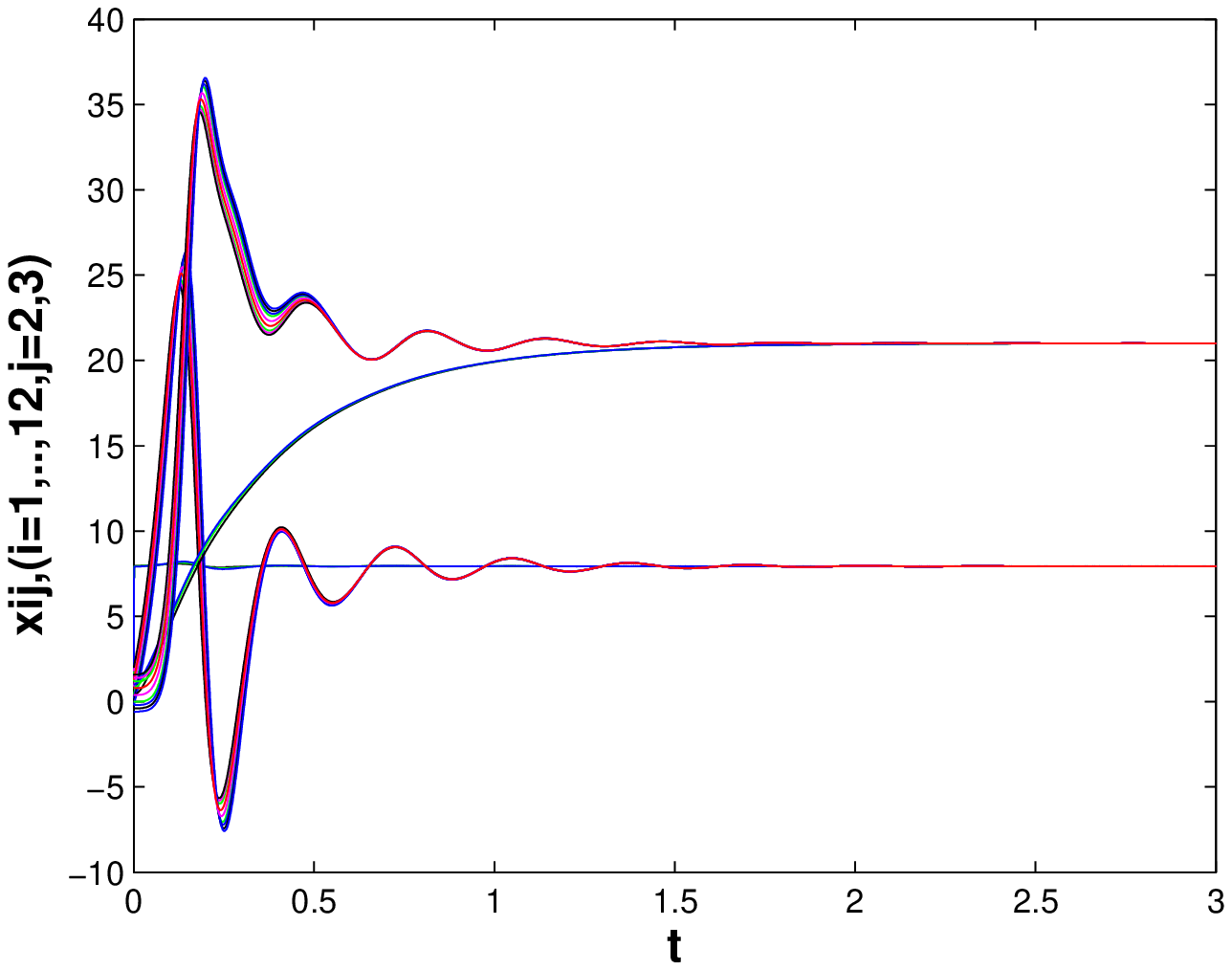}\quad \epsfxsize8cm
\epsfysize5.8cm \epsffile{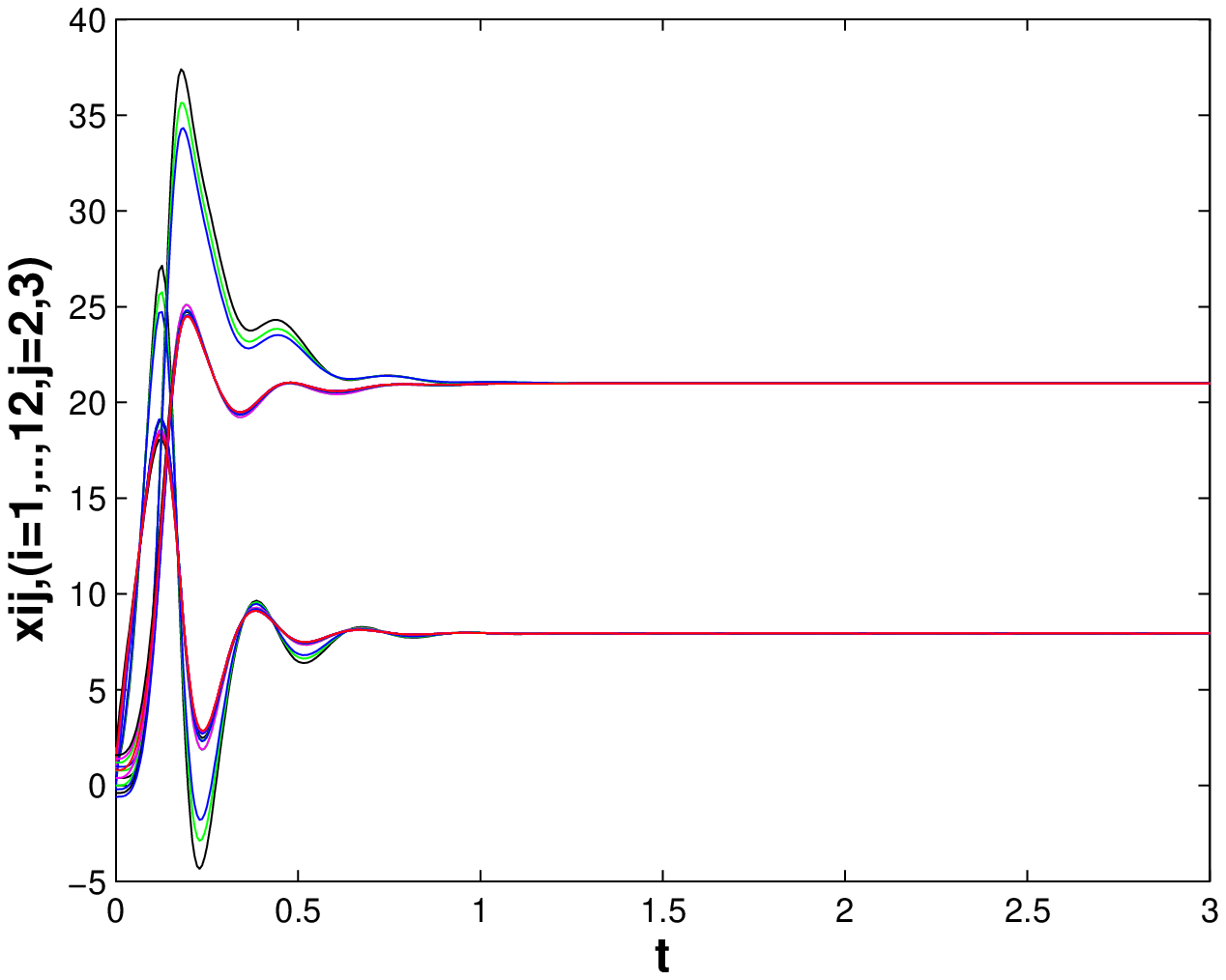}}
\end{center}
\vskip-0.6cm \qquad\qquad\quad\;{\small (a) $CF=c\varepsilon
k=300kc=9000$. \qquad\qquad\qquad\; (b)
$CF=c\varepsilon(\sum_{i=1}^kn_i)=2.5(\sum_{i=1}^kn_i)c=225$.}
 \vskip0.1cm \qquad {\small Fig. 5
\,\,\,\, Synchronization of network \dref{cost6} with coupling
strength $c=10$: (a) pinning $k$ central nodes; (b) pinning all the
non-central nodes.}

From Fig. 5, it can be seen that although all the non-central nodes
are pinned, the cost is lower and the effect is better than the case
where the central nodes are pinned.

By using the Kronecker product, the controlled network \dref{syn5}
can be rewritten as
\begin{equation} \label{cost9}
\dot e(t)=F(x,s)+c(A-G)\otimes\Gamma e(t),
\end{equation}
where $e(t)=(e_1^T(t),\cdots,e_N^T(t))^T$,
$F(x,s)=((f(x_1(t))-f(s(t)))^T,\cdots,(f(x_N(t))-f(s(t)))^T)^T$,
$G=diag\{\underbrace{0,\cdots,0}_k,\underbrace{\varepsilon,\cdots,\varepsilon}_{N-k}\}\in
R^{N\times N}$ and $\Gamma=diag\{r_1,r_2,\cdots,r_n\}$, $r_i=1$ or
$0,\;i=1,\cdots,n$.

\textbf {Theorem 3} If there are a positive diagonal matrix $P$ and
constants $\mu>0,\,\varepsilon>0,\,\tilde{k}>0,\,c>0$, such that
$n_1>\tilde{k}$, $\varepsilon>\tilde{k}-1$,
$\varepsilon>\frac{\tilde{k}(n_1+1-\tilde{k})}{n_1-\tilde{k}}$ and
$$
(x-y)^TP(f(x,t)-f(y,t)-c\tilde{k}\Gamma(x-y))\leq-\mu(x-y)^T(x-y),
$$
then the controlled network \dref{cost7} is globally exponentially
synchronized to $s(t)$.

{\bf Proof:} Choose a Lyapunov function as
$V(t)=\frac{1}{2}e^T(t)(I_N\otimes P)e(t)$. Its time derivative is
$$ \begin{array}{ccl}
\dot V(t)&=&e^T(t)(I_N\otimes P)\dot e(t)\\
         &=&e^T(t)(I_N\otimes P)(F(x,s)+c(A-G)\otimes\Gamma e(t))\\
         &=&e^T(t)(I_N\otimes P)F(x,s)+ce^T(t)((A-G)\otimes P\Gamma)e(t)\\
         &=&e^T(t)(I_N\otimes P)(F(x,s)-c(\tilde{k}I_n\otimes
              \Gamma)e^T(t))+ce^T(t)((A-G)+\tilde{k}I_n)\otimes P\Gamma
              e(t).
\end{array}
$$
By \textrm{Theorem 2}, $(A-G)+\tilde{k}I_n<0$. So,
$$ \begin{array}{ccl}
\dot
V(t)&\leq&-\mu\sum_{i=1}^Ne_i^T(t)e_i(t)\leq-\mu\sum_{i=1}^N\frac{1}{\max\{P_j\}_{1\leq
j\leq n}}e_i^T(t)Pe_i(t)\\
\\
&=&-\frac{\mu}{\max\{P_j\}_{1\leq j\leq n}}e_i^T(t)(I_n\otimes
P)e_i(t)=-\frac{\mu}{\max\{P_j\}_{1\leq j\leq n}}V(t),
\end{array}$$
and the theorem is thus proved. \hfill $\square$

\section{Some non-regular complex networks}

\quad In this section, first, a non-regular coupled network
consisting of 20 nodes is generated following the procedure of the
well-known BA model. Fig. 6 shows the synchronization of the network
with controllers being added to the three ``biggest" nodes of
degrees 15, 13 and 10, respectively. The dynamics of an isolated
node, the inner liking matrix $\Gamma$, and the synchronized state
are the same as that given in Section 2. From Figs. 6(c) and 6(d),
it can be seen that the network does not synchronize any faster,
although a larger feedback gain is used. In Fig. 7, a different
control scheme is applied: eleven nodes with smaller degrees in the
network are pinned with coupling strength $c=6$ and feedback gain
$\varepsilon=5$, yielding cost function $CF=330$. Compared with Fig.
6, a better control performance is obtained with a much smaller cost
function. Figs. 8 and 9 also show that it is more efficient by
pinning nodes with smaller degrees than pinning nodes with larger
degrees. As shown in Fig. 8, to achieve a similar synchronization
effect, a much smaller feedback gain is needed in pinning ``smaller"
nodes than pinning ``larger" ones, thereby the cost function is also
smaller in the former control scheme than that in the latter. Fig. 9
shows that although the coupling strength, the feedback gain and the
cost function are similar, the synchronization effect is better in
pinning ``smaller" nodes than pinning ``larger" ones.
\begin{center}
\quad \unitlength=1cm \hbox{\hspace*{0.1cm} \epsfxsize8cm
\epsfysize5.8cm \epsffile{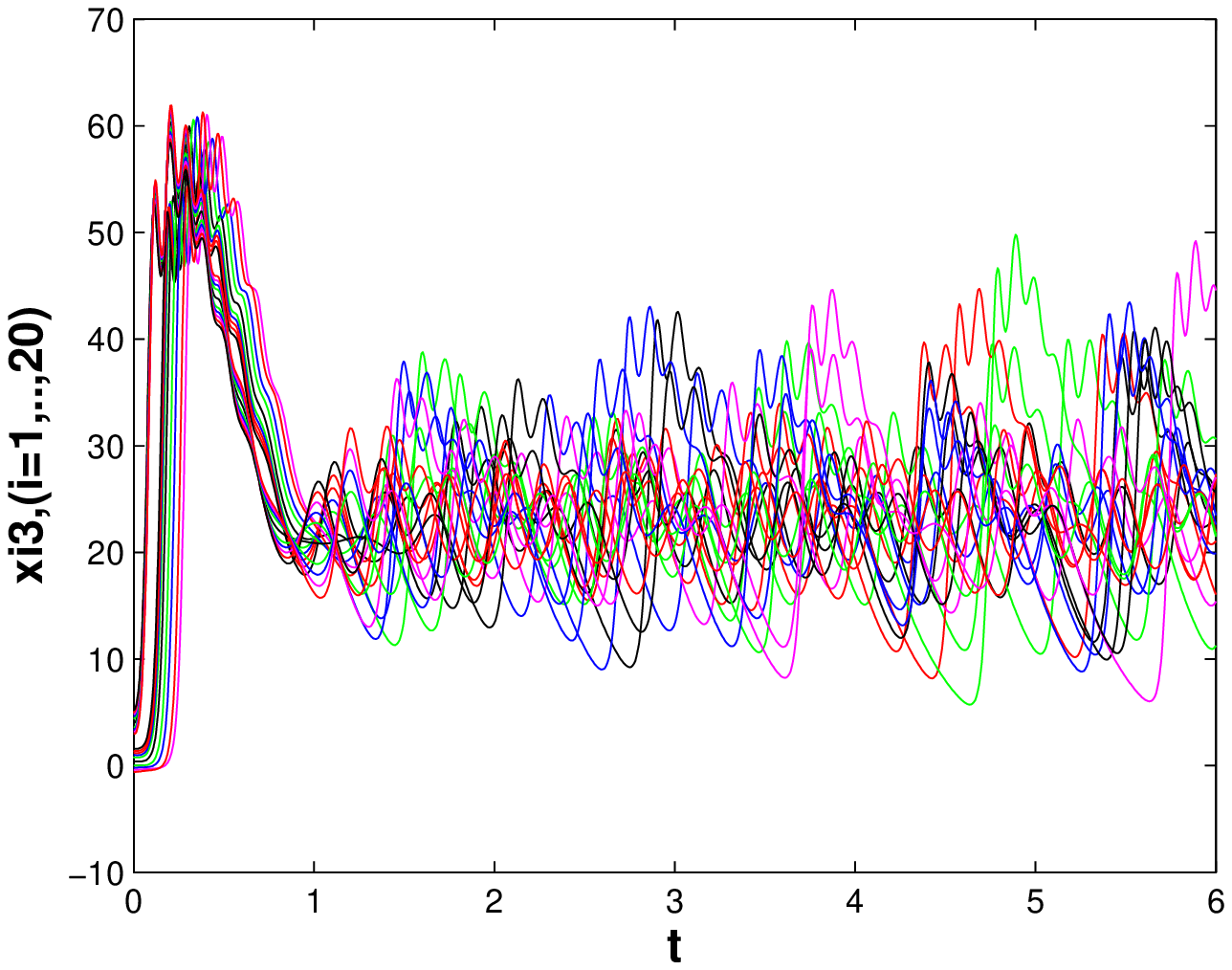}\quad \epsfxsize8cm
\epsfysize5.8cm \epsffile{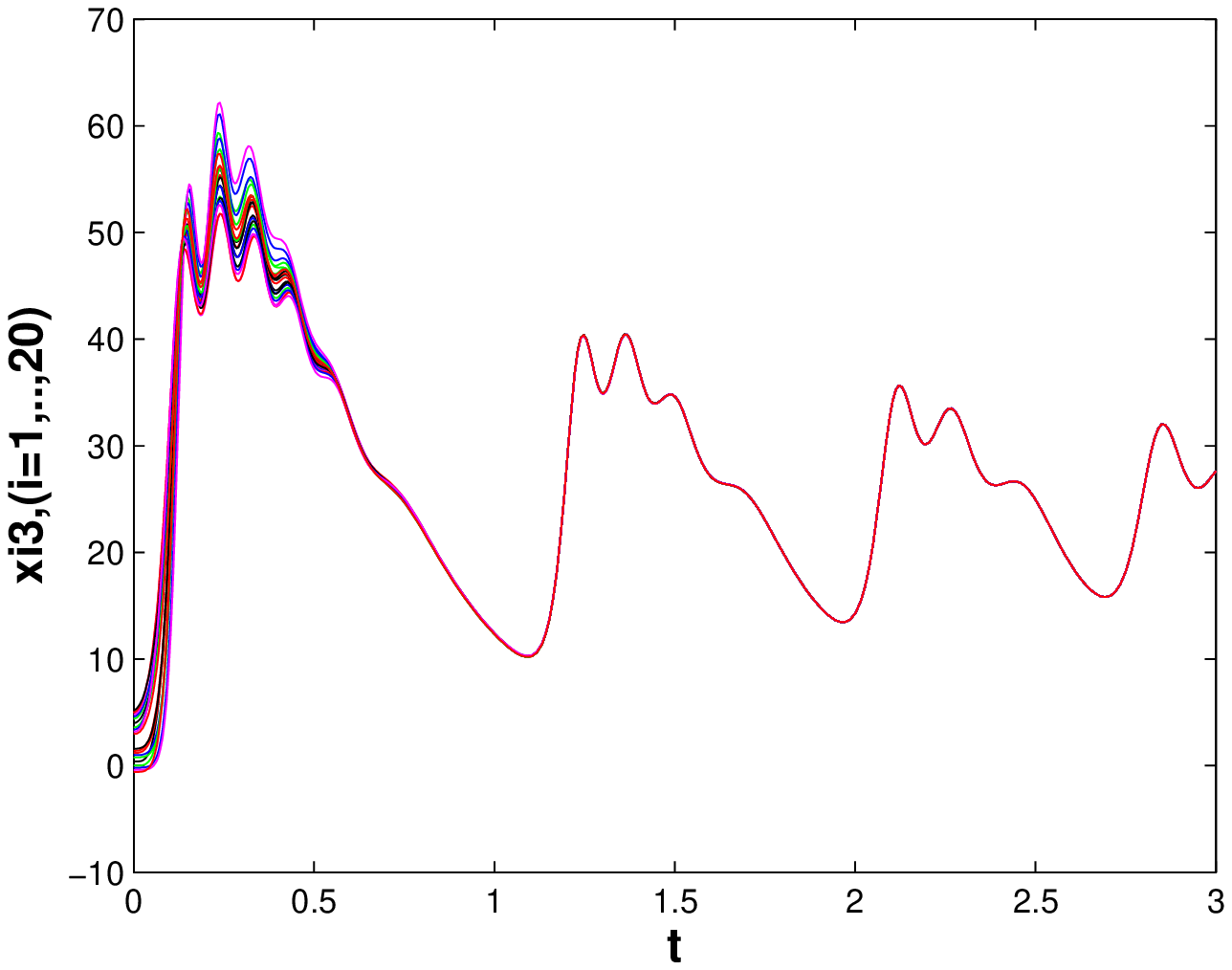}}
\end{center}
\vskip-0.6cm \qquad\qquad\qquad\qquad\qquad{\small (a)
\qquad\qquad\qquad\qquad\qquad\qquad\qquad\qquad\qquad\qquad\qquad\qquad(b)
}
\begin{center}
\quad \unitlength=1cm \hbox{\hspace*{0.1cm} \epsfxsize8cm
\epsfysize5.8cm \epsffile{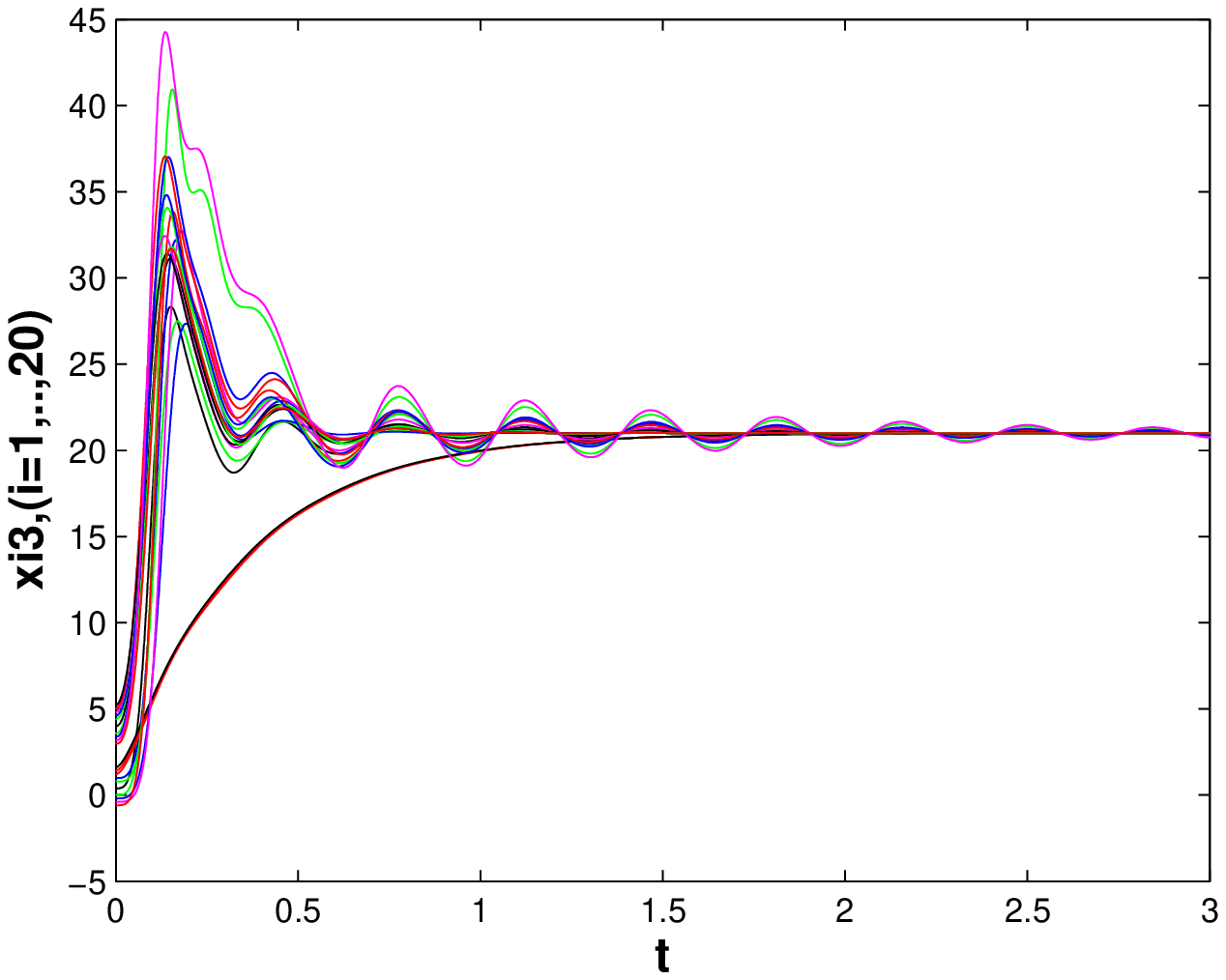}\quad \epsfxsize8cm
\epsfysize5.8cm \epsffile{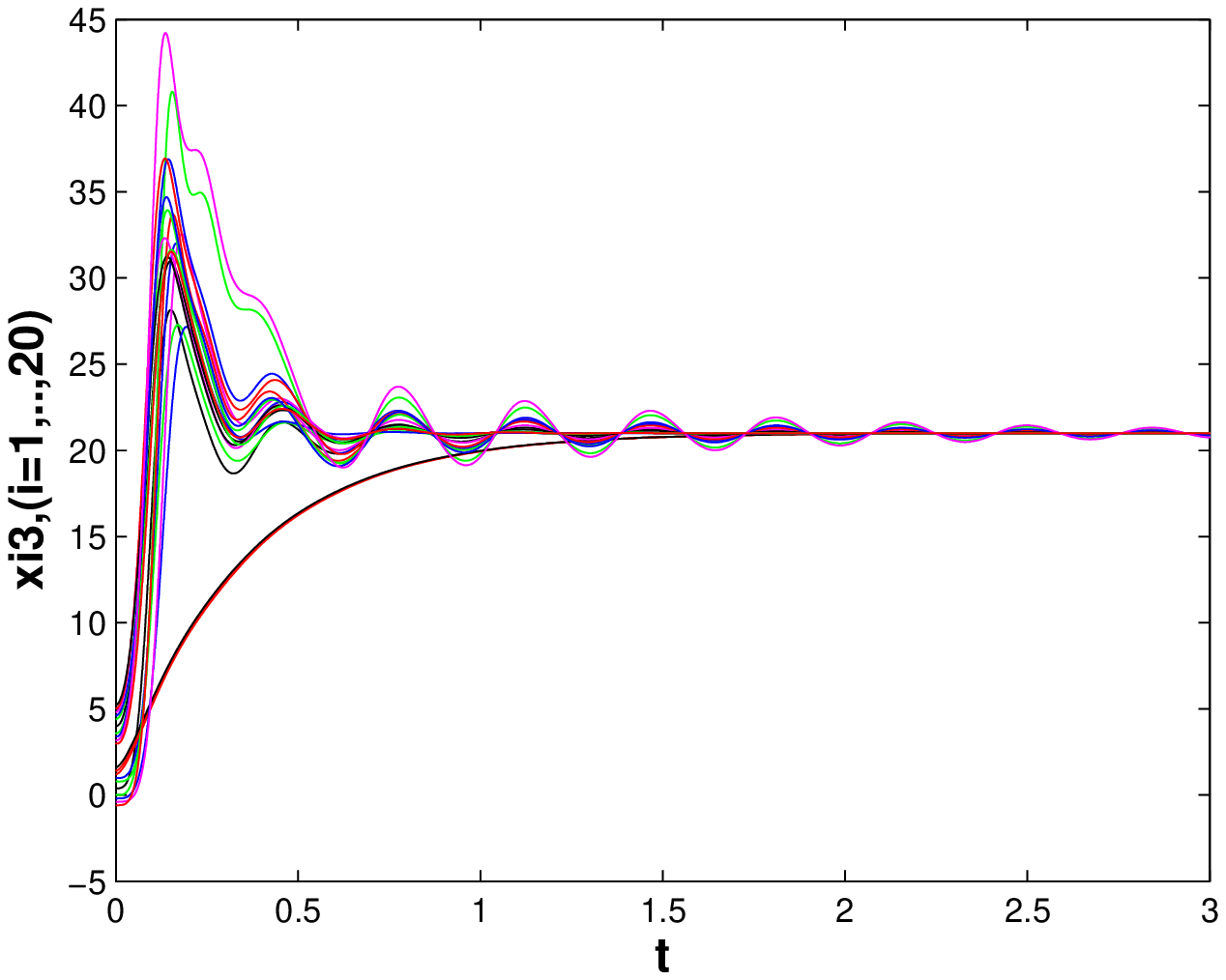}}
\end{center}
\vskip-0.6cm \qquad\qquad\qquad\qquad\qquad{\small (c)
\qquad\qquad\qquad\qquad\qquad\qquad\qquad\qquad\qquad\qquad\qquad\qquad(d)}
\vskip0.1cm \qquad {\small Fig. 6 \,\,\,\, Pinning the three
``biggest" nodes with degrees 15, 13 and 10 in a 20-node non-regular
coupled network: (a) $c=0,\,\varepsilon=0,\,CF=0.$ (b)
$c=6,\,\varepsilon=0,\,CF=0.$ (c) $c=6,\,\varepsilon=500,\,CF=9000.$
(d) $c=6,\,\varepsilon=1000,\,CF=18000.$ }
\begin{center}
\quad \unitlength=1cm \hbox{\hspace*{0.1cm} \epsfxsize8cm
\epsfysize5.8cm \epsffile{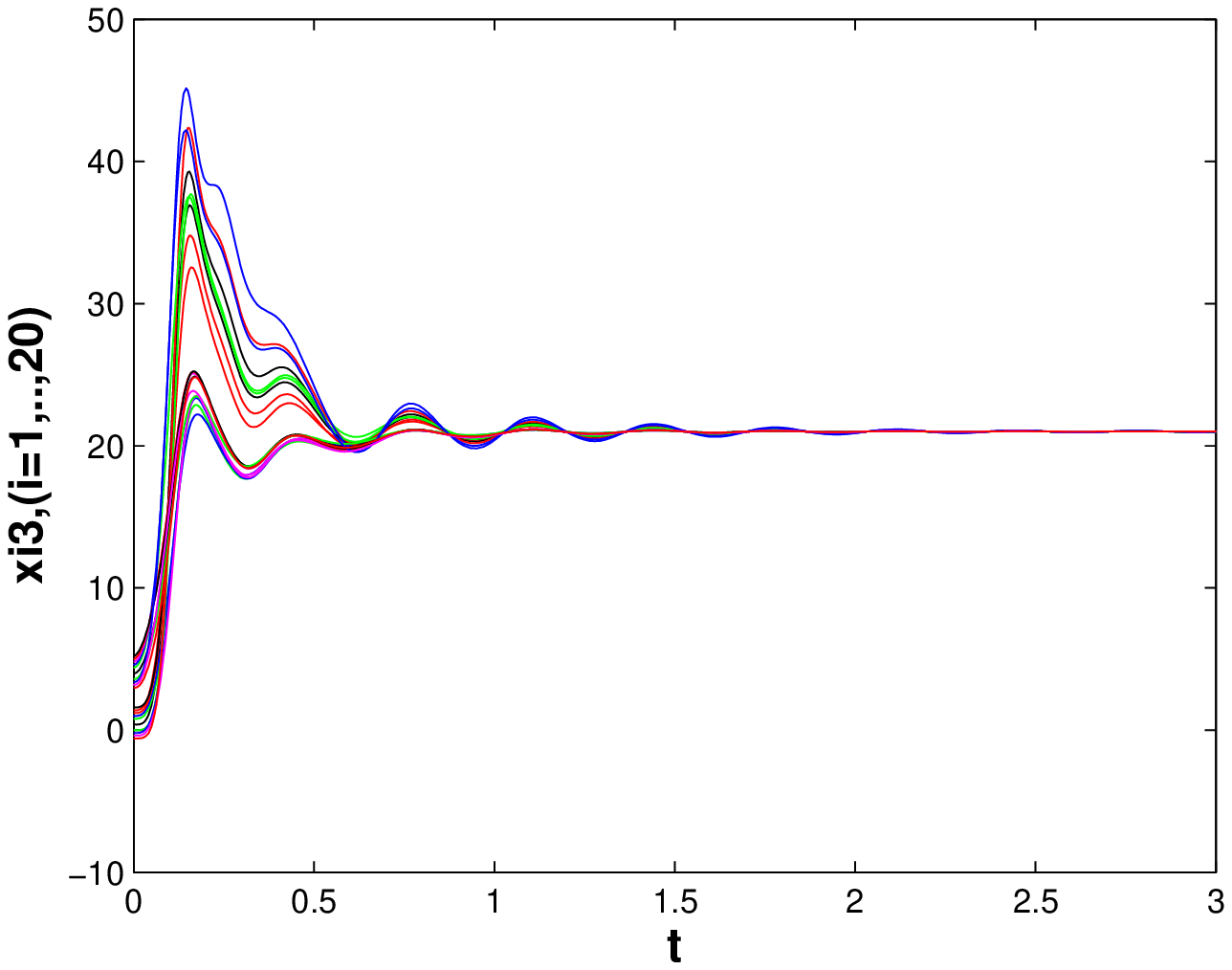}}
\end{center}
\vskip0.1cm \qquad {\small Fig. 7 \,\,\,\, Pinning the eleven
``smaller" nodes in a 20-node non-regular coupled network: $c=6$,
$\varepsilon=5$, $CF=330.$}
\begin{center}
\quad \unitlength=1cm \hbox{\hspace*{0.1cm} \epsfxsize8cm
\epsfysize5.8cm \epsffile{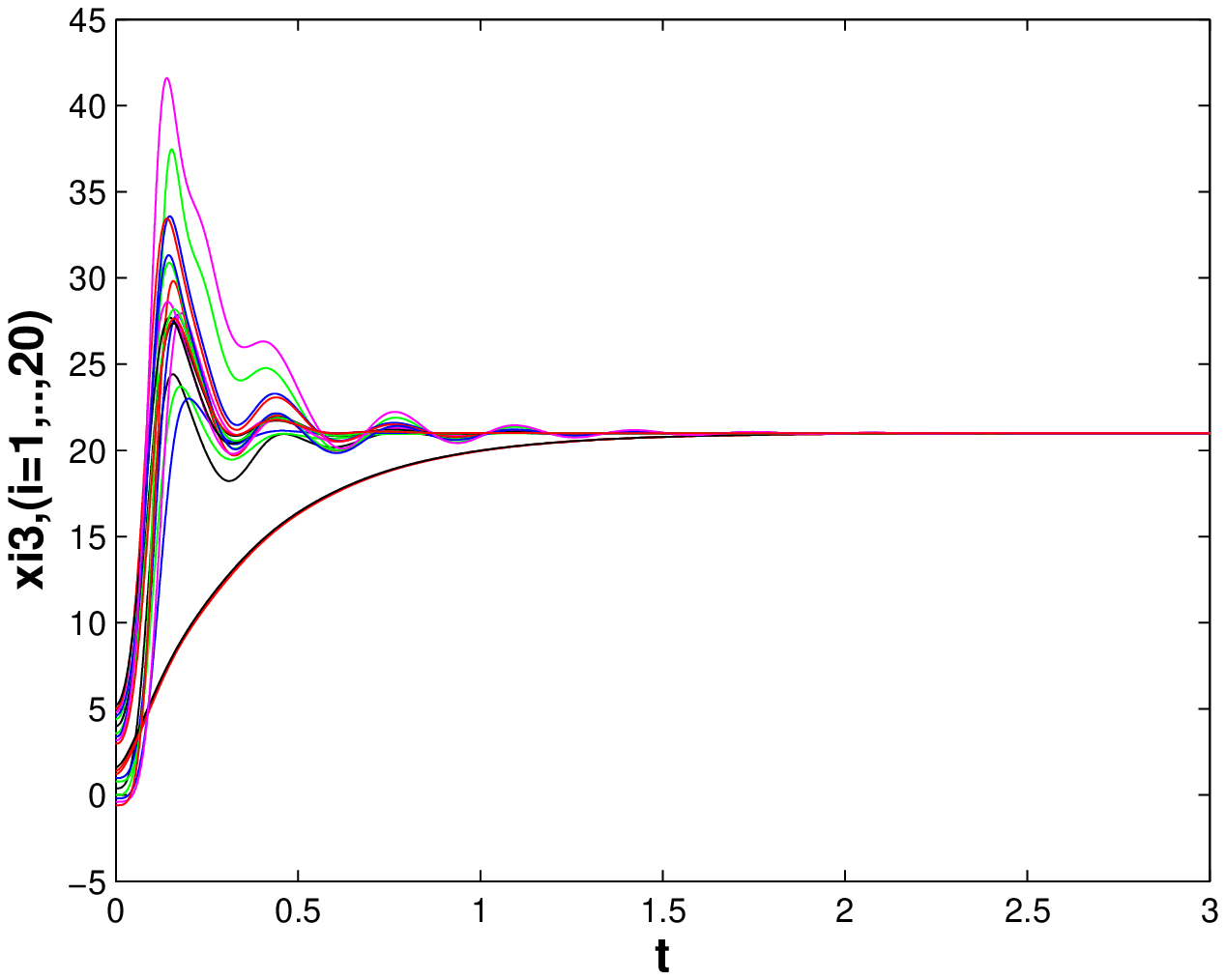}\quad \epsfxsize8cm
\epsfysize5.8cm \epsffile{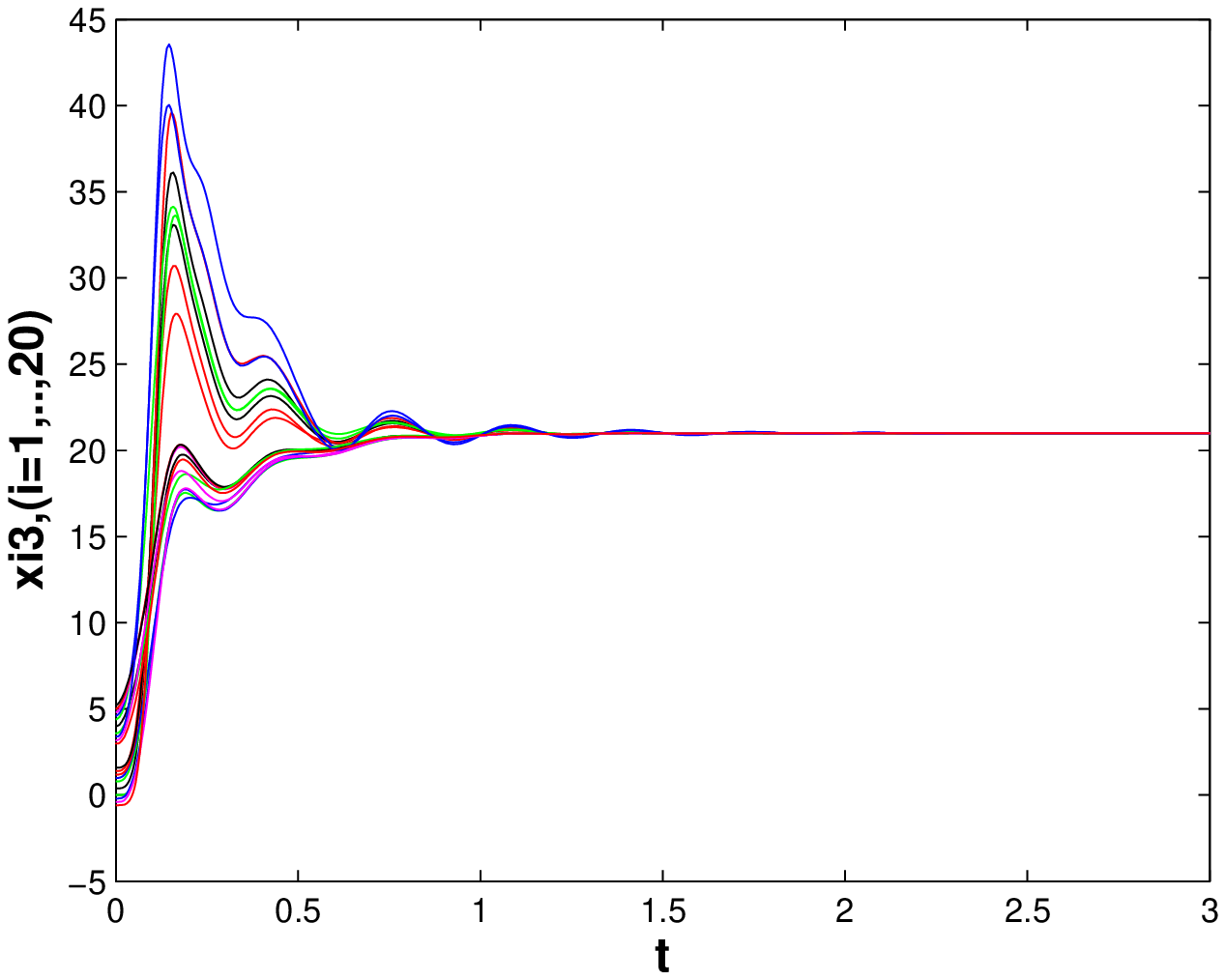}}
\end{center}
\vskip-0.6cm \qquad\qquad\quad\;\;{\small (a)
$c=8,\,\varepsilon=500,\,CF=12000.$
\qquad\qquad\qquad\qquad\qquad\quad (b)
$c=6,\,\varepsilon=8,\,CF=528.$} \vskip0.1cm \qquad {\small Fig. 8
\,(a) Pinning the three ``biggest" nodes with degrees 15, 13 and 10.
(b) Pinning the eleven ``smaller" nodes. }
\begin{center}
\unitlength=1cm \hbox{\hspace*{0.1cm} \epsfxsize5.4cm
\epsfysize4.8cm \epsffile{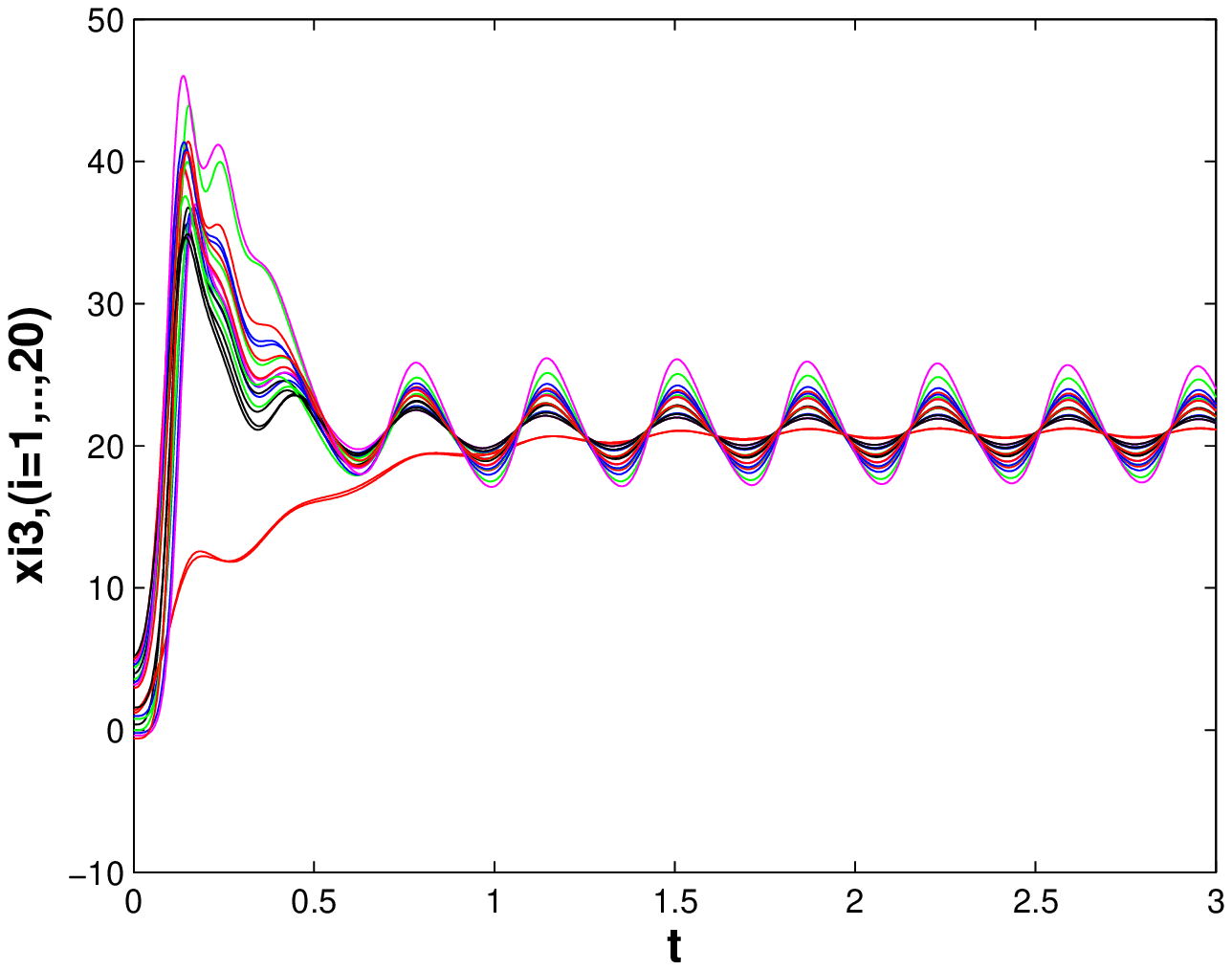}\epsfxsize5.6cm
\epsfysize4.8cm \epsffile{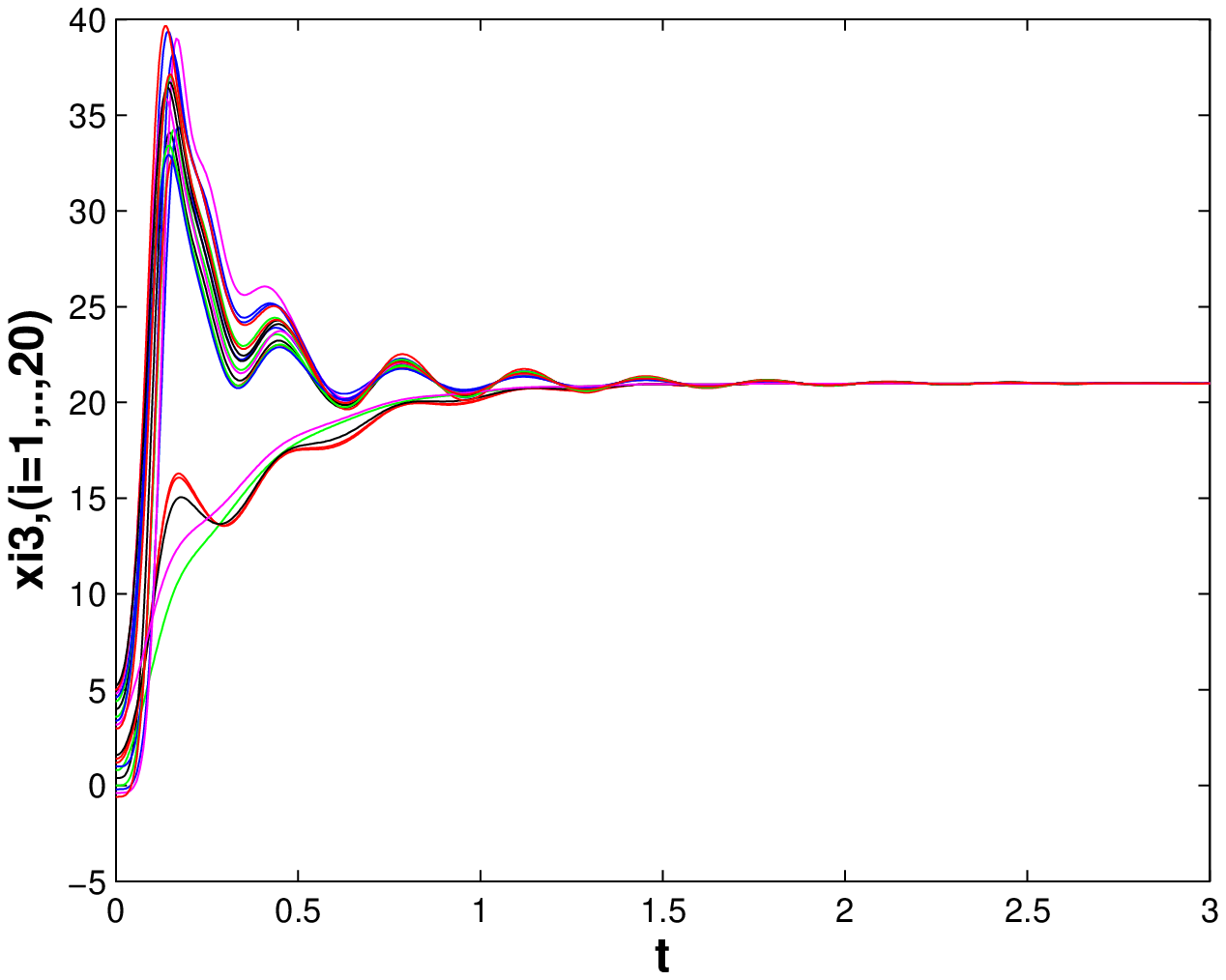}\epsfxsize5.6cm
\epsfysize4.8cm \epsffile{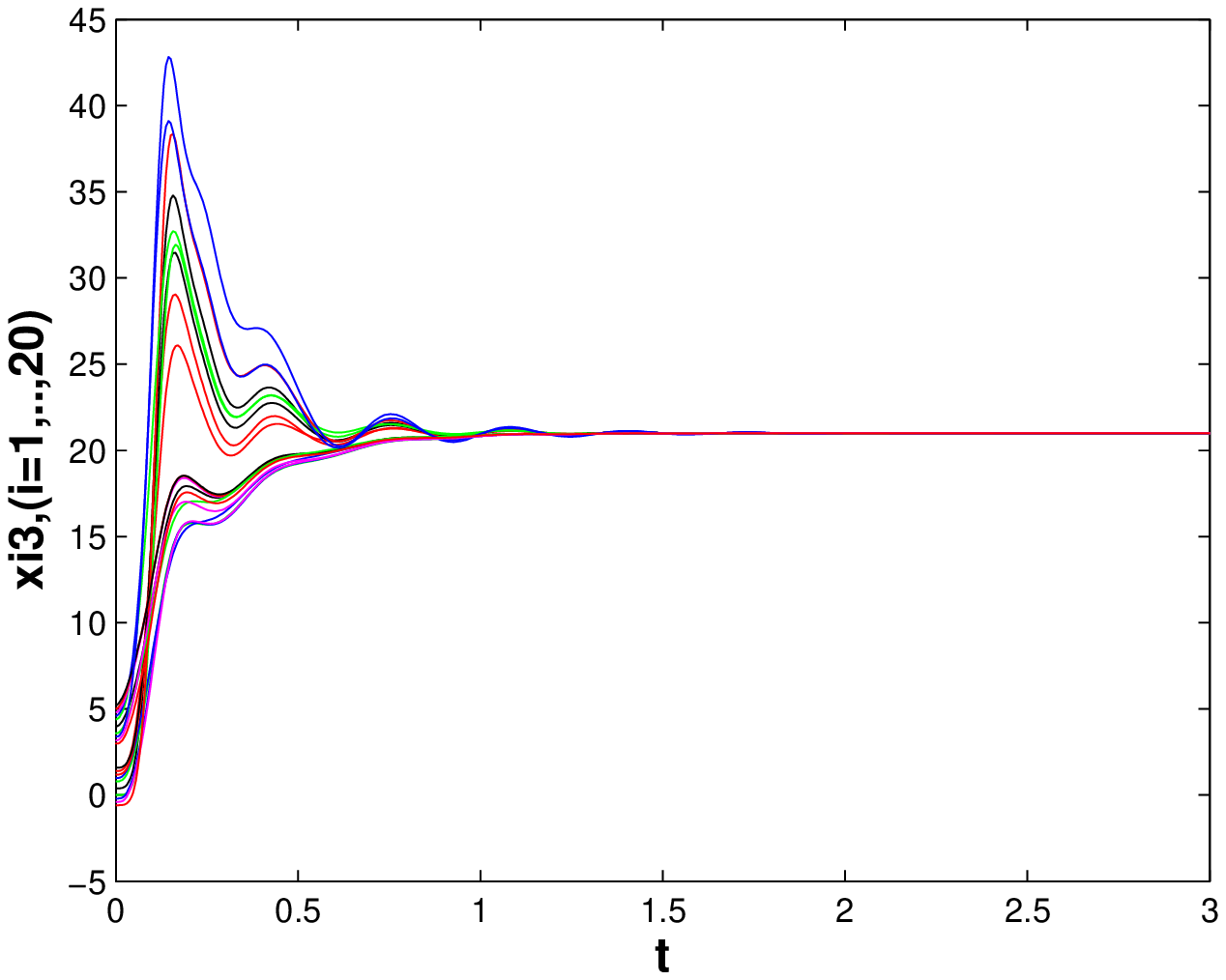}}
\end{center}
\vskip-1cm {\small \qquad\;\,(a) $c=6,\,\varepsilon=55,\,CF=660.$
\qquad\quad (b) $c=6,\,\varepsilon=22,\,CF=660.$ \qquad\quad\, (c)
$c=6,\,\varepsilon=10,\,CF=660.$} \vskip0.1cm {\small\qquad  Fig. 9
\,(a) Pinning the two ``biggest" nodes with degrees 15 and 13. (b)
Pinning the three ``biggest" nodes with degrees 15, 13 and 10, and
the two ``smallest" nodes with the same degrees 3. (c) Pinning the
eleven ``smaller" nodes.}

It seems more efficient to use the control scheme of pinning nodes
with smaller degrees than pinning nodes with larger degrees for
achieving the desired synchronous states of some non-regular coupled
dynamical networks. This is an interesting phenomenon that was not
noticed before.

\section{Conclusions}

\quad In this paper, pinning control to achieve synchronization of
complex dynamical networks is further investigated. In contrary to
the general perception, the nodes with smaller degrees play an
important role in the synchronizability of some networks. It has
been shown by computer simulations in this paper that to achieve a
similar synchronization effect on the networks considered, a smaller
coupling strength, a smaller feedback gain and thus a lower cost
function are needed in the control scheme of pinning nodes with
smaller degrees comparing with those needed in pinning nodes with
larger degrees. In other words, it seems more efficient to use the
control scheme of pinning nodes with smaller degrees than that of
pinning nodes with larger degrees for some dynamical networks to
achieve synchronization, an interesting phenomenon that deserves
further investigation in the future.

%

\end{document}